\documentclass[aps,showpacs,superscriptaddress]{revtex4-2}
\usepackage{amssymb,bm}
\usepackage{graphicx}
\usepackage{amsmath}
\usepackage{epstopdf}
\usepackage{float}
\allowdisplaybreaks
\usepackage[english]{babel}

\begin{document}

\title{Atomic screening and $e^+e^-$ pair photoproduction   at low energies.}

\author{P.A. Krachkov}\email{P.A.Krachkov@inp.nsk.su}
\author{A.I. Milstein}\email{A.I.Milstein@inp.nsk.su}
\affiliation{Budker Institute of Nuclear Physics, SB RAS, Novosibirsk, 630090, Russia}
\affiliation{Novosibirsk State University, 630090 Novosibirsk, Russia}


\begin{abstract}
The effect of screening by atomic electrons on the behavior of electron and positron wave functions in the continuous spectrum at small distances is studied. It is shown that these asymptotics are affected not only by the static potential of atomic electrons, but also by the polarization potential, as well as by the exchange interaction, which is essential for nonrelativistic electrons. A simple analytical expression is obtained for the photoproduction cross section of $e^+e^-$ pair in an atomic field near the threshold. The spectrum and angular distribution of the produced particles are considered. It is shown that screening significantly affects the cross section in the near-threshold region.	
\end{abstract}



\maketitle

\section{Introduction}
At present, the process of $e^+e^-$ pair photoproduction  on an atom has been studied in detail exactly in the parameters of the atomic field at photon energies $\omega\gg m_e$, where $m_e$ is the electron mass, $\hbar=c=1 $. In this case, both the spectra of produced particles and their angular distributions are studied in detail, taking screening into account. The results are obtained using the quasiclassical approximation  with account for the first quasiclassical corrections \cite{LMS2012,KLMUFN16}. The quasiclassical approximation is applicable because  at high energies the main contribution to the cross section is given by large angular momenta of produced particles and small angles between their momenta and the photon momentum. For $\omega\gtrsim m_e$, exact in  $\omega$ results  for the spectra of produced particles are obtained using the exact solution of the Dirac equation in an atomic field \cite{Overbo1968,Overbo1972}. At the same time, there are no results for the angular distribution of produced particles, obtained exactly in the atomic field parameters and $\omega\gtrsim m_e$.

Numerical results in the near-threshold region were obtained in \cite{TsengPratt71,TsengPratt72,TsengPratt79,Overbo1978}.
The analytical  results of \cite{NTS34} obtained for the case of the Coulomb field at $\omega-2m_e\ll m_e$ are cited in many reviews, see e.g. \cite{HGO1980,MKO69}. It has been shown in a recent paper \cite{KLM22} that these results are erroneous both for the spectrum of produced particles and for their angular distribution. At low energies, the main contribution to the pair production cross section comes from small angular momenta of produced particles and small distances $r\sim \lambda_C$, where  $\lambda_C=1/m_e$ is the electron Compton  wavelength. At such distances, the wave functions of non-relativistic electrons with small angular momenta are significantly enhanced by the so-called Sommerfeld-Gamov-Sakharov factor \cite{SGS}, while the wave functions of non-relativistic positrons with small angular momenta are significantly suppressed. Atomic screening strongly affects the behavior of wave functions at large distances of non-relativistic electrons and positrons. The question arises of how screening affects the behavior of the wave functions at small distances, which determine the pair production cross section at low energies. Our work is devoted to the study of this issue. We show that the cross section of the process is significantly affected not only by the static potential of atomic electrons, but also by the polarization potential, as well as by the exchange interaction, which is essential for non-relativistic electrons. We  consider the influence of atomic electrons on the spectrum and angular distribution of produced particles in the process of $e^+e^-$ photoproduction in the atomic field near the threshold.

\section{Wave function of non-relativistic electrons at small distances}\label{vfe}
For the applicability of the non-relativistic approximation, we assume that $Z\alpha\ll 1$ and $v\ll 1$  where $v$ is the electron or positron velocity in a continuous spectrum. We also assume that $Z\alpha/v\sim 1$ and $Z\gg 1$,  where $Z$ is the atomic charge number and $\alpha=1/137$ is the fine structure constant. The latter inequality allows one to use the Thomas-Fermi approximation to describe the static interaction potential $V(r)$ of an electron with an atom (Moli\`ere potential \cite{molier47}),
\begin{align}\label{TF}
	&V(r)= -\dfrac{Z}{r}\,\left[0.1\, e^{-6\,\beta\, r}+0.55\, e^{-1.2\,\beta\, r}+0.35\, e^{-0.3\,\beta\, r}\right]\,,\quad \beta=\dfrac{137}{121}\,Z^{1/3}\,.
\end{align}
In this formula and below  we use the atomic units, in which momentum, distance and energy are measured in  $m_e\alpha$, $1/m_e\alpha$ and $m_e\alpha^2$ , respectively.  The screening radius in these units is  $r_{scr}\sim 1/\beta\sim Z^{-1/3}$, the characteristic size of the ground state is $1/Z$, and the momentum of most atomic electrons is $Z^{2/3}$.

\begin{figure}
\includegraphics[height=6cm]{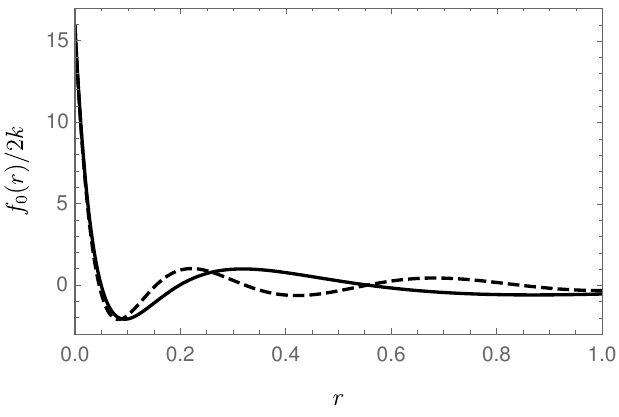}
\caption{Dependence of radial electron wave function $f_{0}(r)$ in units of $2k$ on $r$ in the atomic potential $V(r)$ (solid line) and in the Coulomb attractive potential $-Z/r$ (dashed line)  for $Z=40$ and $k=1$.
 }\label{psi}
\end{figure}

Fig.~\ref{psi} shows the dependence on $r$ of the radial electron wave function $f_l(r)$  in units of $2k$ in the potential \eqref{TF} and in the Coulomb potential $-Z/r$ for the orbital angular momentum $l=0$ , momentum $k=1$, and $Z=40$.

\begin{figure}
	\includegraphics[height=6cm]{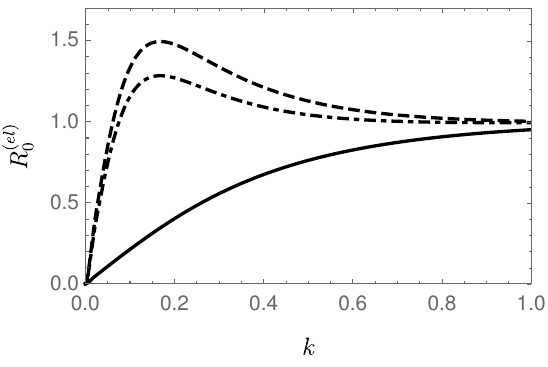}
	\caption{Dependence of $R^{(el)}_0(k)$ on $k$ for $l=0$, $Z=60$ (solid curve), $Z=40$ (dotted curve), and $Z=20$ (dash-dotted curve).}
	\label{psik}
\end{figure}

An analysis of similar plots for different $Z$ shows that for $k\gtrsim 1$ and $r\gtrsim \pi/Z$ the Coulomb functions $f_{0,\,c}(r)$ differ significantly from the functions $f_{0}(r)$ in the atomic potential, while for $r\lesssim \pi/Z$ these functions are almost the same. For $k\ll 1$,  the electron  wave function $f_{0,\,c}(r)$ differ from the functions $f_{0}(r)$  for any $r$. However, for $r\lesssim \pi/Z$ the ratio $f_{0}(r)/f_{0,\,c}(r)$ depends only on $Z$ and $k$ and is independent of $r$. Fig.\ref{psik} shows the dependence of $\displaystyle{R^{(el)}_l(k)=\lim_{r\to 0}[f^2_{l}(r)/f^2_{l ,\,c}(r)}]$ on $k$ for $l=0$ and several values of $Z$, where the functions $f_{0,\,c}(r)$ and $f_{1,\,c}(r)$ for $kr\ll 1$ are \cite{BLP1982}
\begin{align}\label{SZe}
	&f_{0,\,c}(r)=2k\sqrt{C_0^{(el)}(k)}\,,\quad C_0^{(el)}(k)= \dfrac{2\pi\eta_k}{1-e^{-2\pi\eta_k}}\,,\quad \eta_k=\dfrac{Z}{k}\,,\nonumber\\
	&f_{1,\,c}(r)=\dfrac{2k}{3}\,(kr)\sqrt{C_1^{(el)}(k)}\,,\quad C_1^{(el)}(k)= \dfrac{2\pi\eta_k}{1-e^{-2\pi\eta_k}}(1+\eta_k^2)\,.
\end{align}

Note that $f_{0,\,c}(0)/2k$ tends to infinity at $k\to 0$, while $f_{0}(0)/2k$ tends to a nonzero constant, so that $R^{(el)}_0(k)\to 0$ for $k\to 0$. 

\begin{figure}
	\includegraphics[height=6cm]{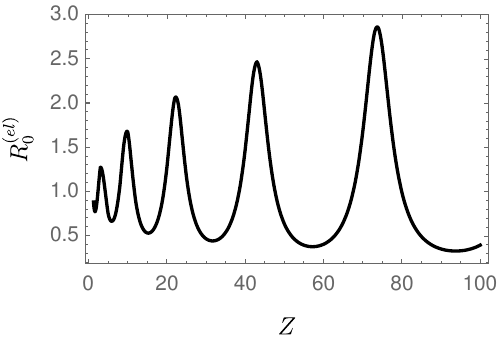}
	\caption{Dependence of $R^{(el)}_0(k)$ on $Z$ for $k=0.2$.}
	\label{psiZ}
\end{figure}

It is interesting that 
$R^{(el)}_0(k)$ depends very strongly on $Z$ for $k\ll 1$, see Fig.\ref{psiZ} which shows the dependence of $R^{(el)}_0(k)$ on $Z$ for $k=0.2$.

\begin{figure}[H]
	\begin{center}
	\includegraphics[height=6cm]{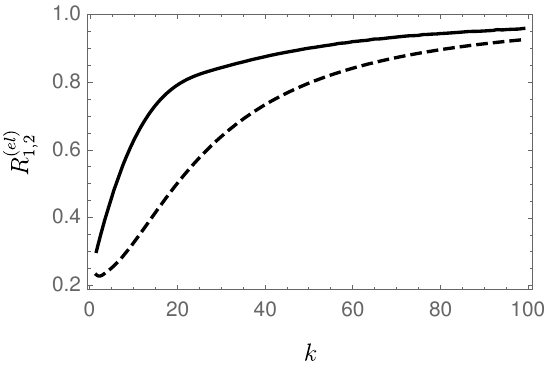}
	\caption{Dependence of $R^{(el)}_l(k)$ on $k$ for $Z=40$, $l=1$ (solid curve) and $l=2$ (dotted curve).}
	\label{R12}
	\end{center}
\end{figure}

Strong oscillations in this figure appear because in the potential \eqref{TF} at certain values of $Z$ there are bound states with a binding energy close to zero. In this case, the scattering length of slow particles becomes much larger than the screening radius, and the characteristic value of the wave function in the region of nonzero atomic potential  differs significantly from the value of the Coulomb wave function. Note that for the angular momentum $l>0$ a significant  deviation of $R^{(el)}_l(k)$ from unity occurs at $k\lesssim Z$  but not at $k\lesssim 1$, as in the case of $l=0$, see Fig.\ref{R12}. This is due to existence of a centrifugal barrier for $l>0$.

For $k\ll 1$, in addition to the static potential $V(r)$, it is also necessary to take into account the van der Waals forces (polarization potential), as well as the exchange interaction, which takes into account the identity of an electron in continuous spectrum   and atomic electrons. Note that the polarization potential essentially depends on $Z$. To illustrate the effect of polarization potential $V_{pol}$, we use the simplest parameterization,
  \begin{align}\label{pol}
  	&V_{pol}(r)= -\dfrac{\alpha_{pol}}{2(r^2+d^2)^2}\,\,,
  \end{align}
where $\alpha_{pol}$ is the static polarizability and $d\sim 1$ is some parameter.
The exchange interaction $V_{ex}$ is taken into account using the model
 \cite{Liberman1968} ,
   \begin{align}\label{ex}
 	&V_{ex}(r)= -\left(\dfrac{3 n_{e}}{\pi}\right)^{1/3}\,\,,\nonumber\\
 	&n_e=\dfrac{Z\beta^2}{4\pi r}\,\left[3.6\, e^{-6\,\beta\, r}+0.792\, e^{-1.2\,\beta\, r}+0.0315\, e^{-0.3\,\beta\, r}\right]\,.
 \end{align}

\begin{figure}[H]
	\begin{center}
	\includegraphics[height=6cm]{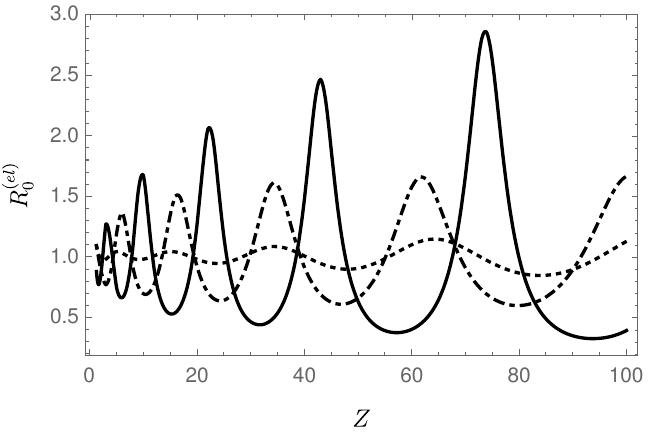}
	\caption{Dependence of $R^{(el)}_0(k)$ on $Z$ at $k=0.2$ and $l=0$ for the potentials $V(r)$ (solid curve),  $V(r) +V_{pol}(r)$ (dash-dotted curve), and  $V(r)+V_{pol}(r)+V_{ex}(r)$ (dashed curve). The parameters $\alpha_{pol}=40$ and $d=2$ are used.}
	\label{AZpol}
	\end{center}
	\end{figure} 

Fig.\ref{AZpol} shows the dependence of $R^{(el)}_0(k)$ on $Z$  with and without account for the exchange and polarization potentials. It is seen that  account for the exchange potential leads to a very small  deviation of $R^{(el)}_0(k)$ from unity.

\section{Wave function of non-relativistic positrons at small distances}\label{vfp}
\begin{figure}[H]
	\begin{center}
	\includegraphics[height=6cm]{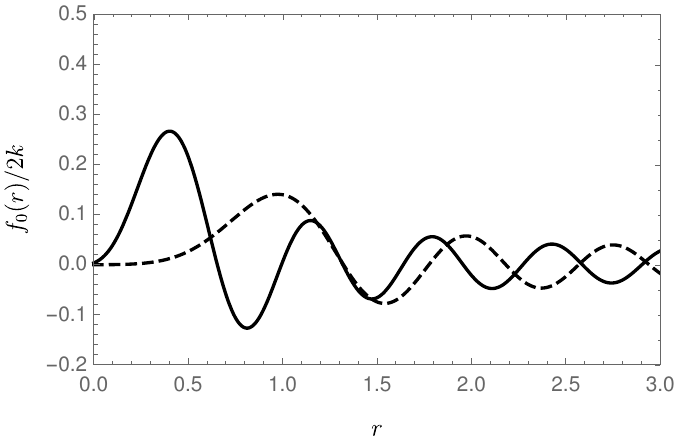}
	\caption{Dependence on $r$  of radial positron wave function $f_{0}(r)$  in the atomic potential $-V(r)$ (solid line) and in the Coulomb repulsive potential $Z/r$ (dashed line)  for $Z=40 $ and $k=10$.}
	\label{vf_pos}
\end{center}
\end{figure}

The wave function of non-relativistic positrons in an atomic field differs significantly from the Coulomb wave function of  positrons for any $k$, see Fig.\ref{vf_pos}.

\begin{figure}[H]
	\begin{center}
	\includegraphics[height=6cm]{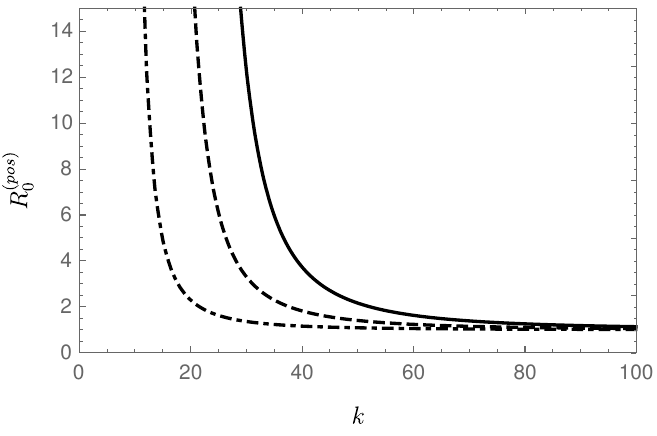}
	\caption{Dependence of $R^{(pos)}_0(k)$ on $k$ for the positron wave functions at $Z=60$ (solid curve), $Z=40$ (dotted curve), and $Z=20$ (dash-dotted curve).}
	\label{Ak_pos}
\end{center}
\end{figure}

This is especially evident at small distances for $k\ll Z$ because of smallness of  positron Coulomb wave function  at $kr\ll 1$,
\begin{align}\label{SZp}
	&f_{0,\,c}(r)=2k\sqrt{C_0^{(pos)}(k)}\,,\quad C_0^{(pos)}(k)= \dfrac{2\pi\eta_k}{e^{2\pi\eta_k}-1}\,,\quad \eta_k=\dfrac{Z}{k}\,,\nonumber\\
	&f_{1,\,c}(r)=\dfrac{2k}{3}\,(kr)\sqrt{C_1^{(pos)}(k)}\,,\quad C_1^{(pos)}(k)= \dfrac{2\pi\eta_k}{e^{2\pi\eta_k}-1}(1+\eta_k^2)\,.
\end{align}
Fig.~\ref{Ak_pos} shows the dependence on $k$ of the ratio  $R^{(pos)}_l(k)=f^2_{l}(0)/f^2_{l,\,c}( 0)$ for the positron wave functions at $l=0$ and several values of $Z$.

The large difference between $R^{(pos)}_0$ and unity begins at $k\lesssim Z$. This is due to  repulsion of the  positron potential at small distances, so that the normalization of the positron wave function in the atomic field is determined by the distances $r\sim 1/\beta$. The properties of wave functions with $l\neq 0$ are similar to the case $l=0$. Note that the influence of the polarization potential on  behavior of the positron wave functions is insignificant. For the case of positron, the exchange interaction is absent, since it is related to  identity of atomic electrons and an incoming particle.

\section{Phenomenological approach to  description of $f_0(0)$ for electron and positron.}\label{fen}

\begin{figure}[H]
	\includegraphics[height=5.2cm]{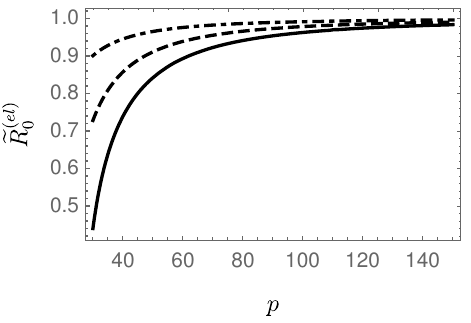}
	\includegraphics[height=5.2cm]{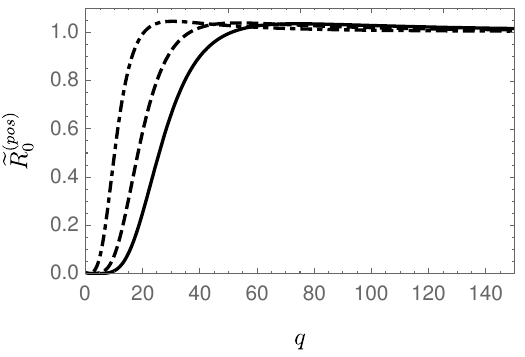}
	\caption{Dependence of the function $\widetilde{R}^{(el)}_0(p)$ on $p$  (left figure) and $\widetilde{R}^{(pos)}_0(q)$ on q (right figure) for $Z=60$ (solid curve), $Z=40$ (dashed curve) and $Z=20$ (dash-dotted curve).}
	\label{RDNep}
\end{figure}

In  Refs.~\cite{TsengPratt72,Overbo1978,TsengPratt79}, a phenomenological approach was proposed to describe $f_0(0)$ in an atomic field. In this approach, in the case of electrons $f_0(0)$ is calculated from $f_{0,\,c}(0)$ by replacing $p\to p_1$ , where $p_1^2/2= p^2/2-V_a$.  In the case of positrons $f_0(0)$ is calculated from $f_{0,\,c}(0)$ by replacing $q\to q_1$,  where $q_1^2/2= q^2/2+V_a$, and $V_a= 1.365\, Z\beta$ is the potential of atomic electrons at small distances.
Naturally, such an approach for electrons can only be used for $p\geq \sqrt{2 V_a}$.
 The dependence of $\widetilde{R}^{(el)}_0(p)$ and $\widetilde{R}^{(pos)}_0(q)$ on $p$ and $q$ is shown in  Fig. \ref{RDNep}. The functions $\widetilde{R}^{(el)}_0(p)$ and $\widetilde{R}^{(pos)}_0(q)$ are obtained from ${R}^{(el)}_0( p)$ and ${R}^{(pos)}_0(q)$ by the  replacement  $p\to p_1$ and $q\to q_1$ in the corresponding functions $f^2_{0,\,c}(0)$.

It is seen that the use of phenomenological approach for electrons and positrons at $k\lesssim Z$ is not justified. For $l\neq 0$ the situation does not change qualitatively.

\section{Cross section of  non-relativistic $e^+e^-$ pair photoproduction in an atomic field.}
Let us discuss the effect of screening on the cross section of $e^+e^-$ pair production in an atomic field by a photon of energy $\omega$ at $\omega-2m_e\ll m_e$. In this section, we use the units $\hbar=c=1$. This cross section, summed over the polarizations of electron and a positron, has the form \cite{BLP1982}
\begin{align}\label{sec0}
	&d\sigma= \dfrac{\alpha pq\varepsilon_p \varepsilon_q\,d\varepsilon_p \,d\Omega_{\bm p}\,d\Omega_{\bm q} }{\omega(2\pi)^4}\,\sum_{\sigma_{1,\,2}=\pm 1}\,|T_{\sigma_1\,,\,\sigma_2}|^2\,,\nonumber\\
	&T_{\sigma_1\,,\,\sigma_2}=\int\,d^3 r\,e^{i\bm k\cdot\bm r} \,\bar U_{\bm p,\,\sigma_1 }^{(out)}(\bm r )\,(\bm\gamma\cdot\bm e_\mu)\,V_{{\bm q},\,\sigma_2}^{(in)}({\bm r})\,,
\end{align}
where $\bm p$ and $\bm q$ are the electron and positron momenta, $\varepsilon_p $ and $\varepsilon_q=(\omega-\varepsilon_p) $ are their energies, $\bm k$ is the photon momentum , $\bm e_{\mu}$ is the photon polarization vector, $\mu$ is its helicity,

 $\bm\gamma$ are the Dirac matrices, $U_{{\bm p},\,\sigma_1} ^{(out)}({\bm r})$ is a positive-frequency solution of the Dirac equation in an atomic field, which at large distances contains a plane wave and a converging spherical wave, 
$V_{{\bm q,\,\sigma_2}}^{(in)}({\bm r})$ is a negative-frequency solution of the Dirac equation in an atomic field, which at large distances contains a plane wave and a diverging spherical wave. The cross section summed over the polarizations of  electron and positron is independent of photon polarization. Therefore, for simplicity of calculations, we chose the circular polarization of photon. Note that for a neutral atom the plane wave is not distorted, in contrast to the case of the Coulomb field.

The wave functions $U_{{\bm p,\,\sigma }}^{(out)}({\bm r})$ have the form \cite{BLP1982}
\begin{align}\label{U}
	& U_{{\bm p,\,\sigma }}^{(out)}({\bm r})=\dfrac{4\pi}{2p}\sum_{l,m}\dfrac{i^lY^*_{l,m-\sigma/2}(\bm n_{\bm p})}{\sqrt{2l+1}}\,\Bigg[ e^{-i\delta_l^{(-)}}\, \sqrt{l+1/2+m\sigma}\,
	\begin{pmatrix}
		f_l^{(-)}\,\Omega_{l+1/2,\,l,\, m}\\
	 g_l^{(-)}\Omega_{l+1/2,\,l+1,\,m}
\end{pmatrix}\nonumber\\
&-\sigma e^{-i\delta_l^{(+)}}\, \sqrt{l+1/2-m\sigma}\,
\begin{pmatrix}
	f_l^{(+)}\,\Omega_{l-1/2,\,l,\, m}\\
	- g_l^{(+)}\Omega_{l-1/2,\,l-1,\,m}
\end{pmatrix}	
	\,\Bigg]\,.
\end{align}
Here $f_l^{(\pm)}(r)$ and $g_l^{(\pm)}(r)$ are the radial components of the positive-frequency wave function, $Y_{l,M}(\bm n )$ are spherical functions, and $\Omega_{j,l,m}(\bm n)$ are spherical spinors. The asymptotics of $f_l^{(\pm)}(r)$ and $g_l^{(\pm)}(r)$ at large distances are
\begin{align}
	&
\begin{pmatrix}
	f_l^{(\pm)}\\
	g_l^{(\pm)}
\end{pmatrix}\,  \underset{r\to\infty}{\to} \dfrac{2}{r\sqrt{2\varepsilon_{p}}}
\begin{pmatrix}
\sqrt{\varepsilon_{p}+m_e}\,\sin(pr-l\pi/2+\delta_l^{(\pm)})\\
 \sqrt{\varepsilon_{p}-m_e}\,\cos(pr-l\pi/2+\delta_l^{(\pm)})
\end{pmatrix}\,.
\end{align}
Thus, the  phases $\delta_l^{(\pm)}$ are determined by the behavior of solutions at large distances, and the potential of atomic electrons makes a significant contribution to these phases.

 The wave functions $V_{{\bm q,\,\sigma }}^{(in)}({\bm r})$ have the form
 \begin{align}\label{V}
	& V_{{\bm q,\,\sigma }}^{(in)}({\bm r})=\dfrac{4\pi}{2q}\sum_{L,m}\dfrac{i^{-L}Y^*_{L,m-\sigma/2}(\bm n_{\bm q})}{\sqrt{2L+1}}\,\Bigg[ e^{i\Delta_L^{(-)}}\, \sqrt{L+1/2+m\sigma}\,
	\begin{pmatrix}
		G_L^{(-)}\,\Omega_{L+1/2,\,L+1,\, m}\\
		-F_L^{(-)}\Omega_{L+1/2,\,L,\,m}
	\end{pmatrix}\nonumber\\
	&+\sigma e^{i\Delta_L^{(+)}}\, \sqrt{L+1/2-m\sigma}\,
	\begin{pmatrix}
		G_L^{(+)}\,\Omega_{L-1/2,\,L-1,\, m}\\
		 F_L^{(+)}\Omega_{L-1/2,\,L,\,m}
	\end{pmatrix}	
	\,\Bigg]\,.
\end{align}
Here $F_L^{(\pm)}(r)$ and $G_L^{(\pm)}(r)$ are the radial components of the negative-frequency wave function, which have asymptotics at large distances
\begin{align}
	&
	\begin{pmatrix}
		F_L^{(\pm)}\\
		G_L^{(\pm)}
	\end{pmatrix}\, \underset{r\to\infty}{\to} \dfrac{2}{r\sqrt{2\varepsilon_{q}}}
	\begin{pmatrix}
		\sqrt{\varepsilon_{q}+m_e}\,\sin(qr-L\pi/2+\delta_L^{(\pm)})\\
		-\sqrt{\varepsilon_{q}-m_e}\,\cos(qr-L\pi/2+\delta_L^{(\pm)})
	\end{pmatrix}\,.
\end{align}

The main contribution to the integral over $\bm r$ in  \eqref{sec0} comes from distances  $r\sim\lambda_C$, so that for a non-relativistic electron and positron $ pr\ll 1,\,qr\ll 1$. As a result, the photoproduction cross section of a non-relativistic $e^+e^-$ pair is determined by the orbital angular momenta $l=0,\,1$ and $L=0,\,1$. In this case, it is necessary to take into account all the components of the wave functions, $f_l^{(\pm)}$,
$g_l^{(\pm)}$, $F_L^{(\pm)}$, and $G_L^{(\pm)}$, which can be expanded in $pr$, $qr$, and $Z\alpha$. In addition, in the wave functions for $j=1/2$ it is necessary to make the replacement $(pr)^{\gamma-1}\to 1-(Z\alpha)^2\ln(pr)/2$, where $ \gamma=\sqrt{1-(Z\alpha)^2}$. As for the phases $\delta^{(\pm)}_l$ and $\Delta^{(\pm)}_L$, they can be considered in the non-relativistic approximation, in which $\delta^{(+)}_l =\delta^{(-)}_l$ and $\Delta^{(+)}_L=\Delta^{(-)}_L$. As a result, the differential cross section  is independent of the phases $\delta^{(\pm)}_l$ and $\Delta^{(\pm)}_L$:
\begin{align}\label{KM}
&	d\sigma=\dfrac{\alpha(Z\alpha)^2pq\,d\Omega_{\bm p}\,d\Omega_{\bm q}\,d\varepsilon_p}{16(2\pi)^2m_e^5}\,\Bigg\{C_0^{(el)}(p)C_0^{(pos)}(q)R_0^{(el)}(p)R_0^{(pos)}(q)\,\dfrac{9\pi^2}{16}(Z\alpha)^2
\nonumber\\
&+\dfrac{p^2}{m_e^2}C_1^{(el)}(p)C_0^{(pos)}(q)R_1^{(el)}(p)R_0^{(pos)}(q)\,\sin^2\theta_p
\nonumber\\
&+\dfrac{q^2}{m_e^2}C_0^{(el)}(p)C_1^{(pos)}(q)R_0^{(el)}(p)R_1^{(pos)}(q)\,\sin^2\theta_q\Bigg\}\,.
\end{align}
Here $C_{0,1}^{(el)}(p)$ and $C_{0,1}^{(pos)}(q)$ are defined in \eqref{SZe} and \eqref{SZp}, $\eta_k= mZ\alpha/k$, $\theta_p$ is the angle between vectors $\bm p$ and $\bm k$, $\theta_q$ is the angle between vectors $\bm q$ and $\bm k$, the functions $R^{(el)}_l(p)$ and $R^{(pos)}_L(q)$ are discussed  in previous sections. Integrating \eqref{KM} over the angles of vectors $\bm p$ and $\bm q$, we obtain

\begin{align}\label{KM1}
	&	d\sigma=\dfrac{\alpha(Z\alpha)^2pq\,d\varepsilon_p }{4m_e^5}\,\Bigg\{C_0^{(el)}(p)C_0^{(pos)}(q)R_0^{(el)}(p)R_0^{(pos)}(q)\,\dfrac{9\pi^2}{16}(Z\alpha)^2
	\nonumber\\
	&+\dfrac{2p^2}{3m_e^2}C_1^{(el)}(p)C_0^{(pos)}(q)R_1^{(el)}(p)R_0^{(pos)}(q)+\dfrac{2q^2}{3m_e^2}C_0^{(el)}(p)C_1^{(pos)}(q)R_0^{(el)}(p)R_1^{(pos)}(q)\Bigg\}\,.
\end{align}
Note that the obtained results \eqref{KM} and \eqref{KM1} are universal and  independent of the explicit form of the atomic potential.

\begin{figure}[H]
	\begin{center}
	\includegraphics[height=6cm]{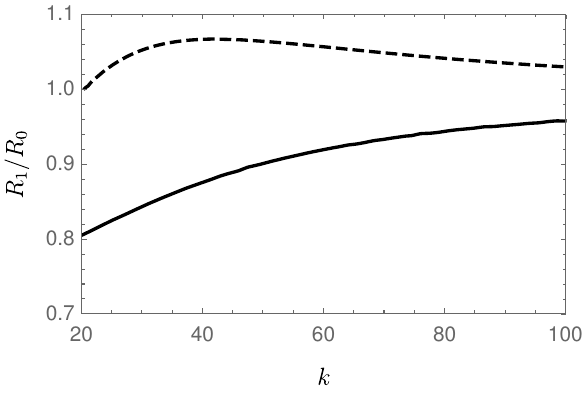}
		\caption{Dependence on $k$ of functions $R^{(el)}_1(k)/R^{(el)}_0(k)$ (solid curve) and $R^{(pos)}_1(k)/R ^{(pos)}_0(k)$ (dashed curve) for $Z=40$.}
	\label{R1R0}
\end{center}
\end{figure}

The case of the Coulomb field corresponds to the substitution $R_l^{(el)}(p),\,R_l^{(pos)}(q)\to 1$. After that the results coincide with that of \cite{KLM22}, in which the errors made in the well-known work \cite{NTS34} were corrected.
Since the dependence of $R^{(el)}_0(p)$ on $p$ differs from the dependence of $R^{(el)}_1(p)$, as well as the dependence on $q$ of  $R^{ (pos)}_0(q)$ and $R^{(pos)}_1(q)$, then the angular dependence  of the cross section $d\sigma$ in the atomic field does not coincide with the angular dependence  of the cross section in the Coulomb field. Fig.~\ref{R1R0} shows the dependence on $k$ of the ratios $R^{(el)}_1(k)/R^{(el)}_0(k)$ and $R^{(pos)}_1(k)/R^{(pos)}_0(k)$.

\section{Conclusion}
In this paper, we study the influence of screening  on the behavior of electron and positron non-relativistic wave functions in the continuous spectrum at small distances. It is shown that  the asymptotic behavior of the electron wave functions is significantly affected not only by the static potential of atomic electrons, but also by the polarization potential and the exchange interaction.
For positrons, the effect of polarization potential on the behavior of  wave functions is not significant. A simple analytical formula is obtained for the differential cross section of $e^+e^-$ pair photoproduction  in an atomic field in the near-threshold region. The results \eqref{KM} and \eqref{KM1} for this cross section  are universal and  independent of the explicit form of the atomic potential.
It is shown that screening significantly affects the cross section.

 \end{document}